\documentclass[reprint,superscriptaddress,amsmath,amssymb,aps,nofootinbib]{revtex4-1}
\usepackage{graphicx}% Include figure files
\usepackage[colorlinks,linkcolor=blue,citecolor=blue,urlcolor = blue]{hyperref}% add hypertext capabilities
\usepackage{multirow} %table

\begin{document}

\title{Polymer Network Diffusion in Charged Gels}

\author{Shoei~Sano}
\affiliation{Department of Bioengineering, The University of Tokyo, Hongo, Bunkyo-ku, Tokyo 113-8656, Japan.}
\author{Takashi~Yasuda}
\affiliation{Department of Bioengineering, The University of Tokyo, Hongo, Bunkyo-ku, Tokyo 113-8656, Japan.}
\author{Takeshi~Fujiyabu}
\affiliation{Department of Bioengineering, The University of Tokyo, Hongo, Bunkyo-ku, Tokyo 113-8656, Japan.}
\author{Naoyuki~Sakumichi}
\email[Correspondence should be addressed to N.~Sakumichi or T.~Sakai: ]
{sakumichi@gel.t.u-tokyo.ac.jp}
\affiliation{Department of Chemistry and Biotechnology, The University of Tokyo, Hongo, Bunkyo-ku, Tokyo 113-8656, Japan.}
\author{Takamasa~Sakai}
\email{sakai@gel.t.u-tokyo.ac.jp}
\affiliation{Department of Chemistry and Biotechnology, The University of Tokyo, Hongo, Bunkyo-ku, Tokyo 113-8656, Japan.}
\date{\today}

\begin{abstract}
The swelling kinetics of charged polymer gels reflect the complex competition among elastic, mixing, and ionic contributions. 
Here, we used dynamic light scattering to investigate the collective diffusion coefficient of model gels, whose polymer network structure was controlled so that the three contributions were comparable. 
We demonstrate that the collective diffusion coefficient stems from the sum of elastic, mixing, and ionic contributions, without evident cross-correlations. 
The significant ionic contribution conforms to the Donnan equilibrium, which explains equilibrium electrical potential gradients in biological systems.
\end{abstract}

\maketitle

Charged gels, characterized by ionic species immobilized in the polymer network, exhibit enhanced swelling compared to electrically neutral gels because of the excess osmotic pressure caused by the Donnan effect \cite{donnan1911theory,adair1923donnan}. 
Their remarkable ability to swell up to $1000$ times their dry weight has facilitated their use as superabsorbent polymers for diapers, horticultural water retention agents, and self-healing concrete.
This has led to extensive investigation of their swelling behaviors \cite{schneider2004discontinuous,cheng2017preparation,ricka1984swelling,jeon1998swelling,tang2020swelling}.

Swelling kinetics are crucial for practical applications of these gels. 
Tanaka and co-workers \cite{tanaka1973spectrum,tanaka1979kinetics} modeled the swelling kinetics of electrically neutral polymer gels, introducing the collective diffusion coefficient $D$ of a polymer network as
\begin{equation}
D=\frac{K+\frac{4}{3}G}{f},
\label{eq:THB}
\end{equation}
where $K$ is the osmotic bulk modulus, $G$ is the shear modulus, and $f$ is the friction coefficient (per unit volume) between polymer networks and solvents.
However, quantifying $D$ based on Eq.~(\ref{eq:THB}) remains challenging~\cite{munch1977inelastic} due to difficulties in estimating $K$ and $f$. 
Moreover, $G \ll K$ is often assumed, leading to $D = K/f$ \cite{tanaka1979kinetics}, even though this is an oversimplification.

Equation~(\ref{eq:THB}) extends to charged gels by including ionic contribution to $K$ from Donnan equilibrium \cite{donnan1911theory,ricka1984swelling,adair1923donnan}. 
However, validation of this approach is complicated due to the complex competition among polymer-solvent mixing, elastic, and ionic contributions. 
To date, only a few scaling laws between $D$ and the polymer volume fraction $\phi$ have been experimentally examined, including $D \sim \phi^{1/2}$ at low salt concentrations \cite{skouri1995swelling,joosten1991dynamic,raasmark2005fast} and $D \sim \phi^{2/3}$ at high salt concentrations \cite{morozova2017elasticity}. 
The main obstacle in experimental validation arises from the difficulty in controlling polymer networks. 
Consequently, studies often use natural polymers like xanthan, chitosan, and hyaluronic acid, which lack precise control over network structure, have been typically used, limiting accurate gel dynamics research.

\begin{figure}[b!]
\centering
\includegraphics[width=\linewidth]{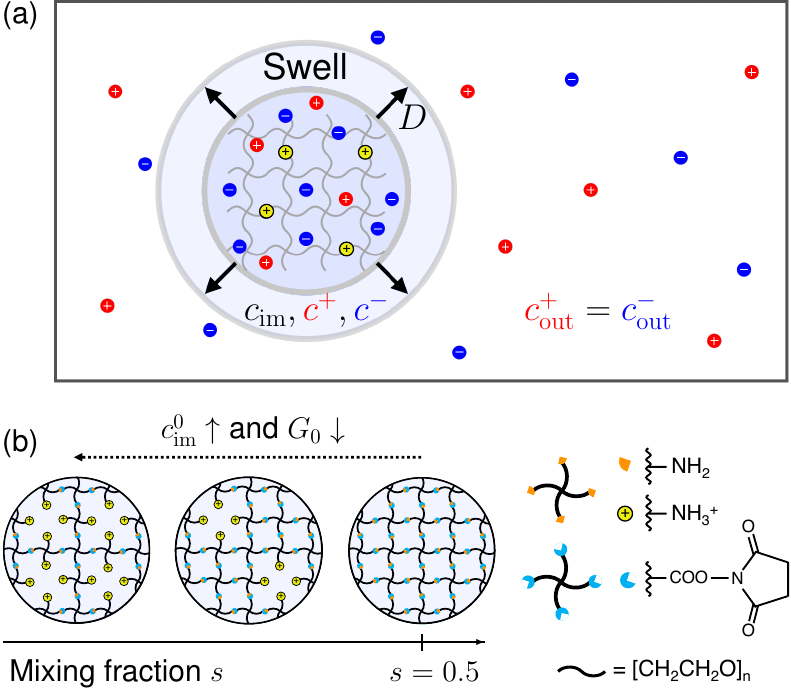}
\caption{
(a) Schematic illustration of the swelling of a charged gel. 
The charged gel with immobilized charge $c_\mathrm{im}$ is equilibrated in the outer salt solution, whose mobile ion $c_\mathrm{out}$ is equivalent to that of cations and anions: $c_\mathrm{out} = c_\mathrm{out}^{+} = c_\mathrm{out}^{-}$.
The mobile cations and anions inside the gels ($c^{+}$ and $c^{-}$) follow the Donnan equilibrium \cite{donnan1911theory,ricka1984swelling,adair1923donnan}: $c^+ c^- = (c_\mathrm{out})^2$.
(b) Schematic illustration of charged gels with precisely controlled polymer networks.
The unreacted end functional groups yield the immobilized charge in the gel at the as-prepared state $c_\mathrm{im}^0$ with the decrease of the shear modulus $G_0$, which is controlled by nonstoichiometrically mixing two types of precursors in an \textit{AB}-type polymerization system.
}
\label{fig:schematic}
\end{figure}

Recently, we have developed a strategy to synthesize tunable model polymer networks \cite{sakai2008design} and identified the equations governing gel elasticity, including negative energy elasticity \cite{yoshikawa2021negative,sakumichi2021linear}, as well as the equation of state of osmotic pressure \cite{yasuda2020universal,sakumichi2022semidilute}.
We have also quantitatively validated Eq.~(\ref{eq:THB}) for neutral gels \cite{fujiyabu2018three,fujiyabu2019shear,kim2020mixing,fujiyabu2021temperature}, by decomposing $D$ into mixing ($D_\mathrm{mix}$) and elastic ($D_\mathrm{el}$) contributions as
$D=D_\mathrm{mix}+D_\mathrm{el}$,
where $D_\mathrm{mix}$ and $D_\mathrm{el}$ are expressed by measurable macroscopic properties.

This Letter extends the relationship $D=D_\mathrm{mix}+D_\mathrm{el}$ to charged gels, introducing an ionic contribution $D_\mathrm{ion}$:
\begin{equation}
D=D_\mathrm{mix}+ D_\mathrm{el}+ D_\mathrm{ion}.
\label{eq:D-ion-gel}
\end{equation}
To validate Eq.~(\ref{eq:D-ion-gel}), we experimentally investigated $D$ in charged gels with comparable $D_\mathrm{mix}$, $D_\mathrm{el}$, and $D_\mathrm{ion}$ via dynamic light scattering (DLS).
We controlled the molar concentrations of immobilized charge in the gel at the as-prepared state $c_\mathrm{im}^0$ and equilibrated the gels in various outer salt solutions with diverse molar concentrations of mobile ions $c_\mathrm{out}$ [Fig.~\ref{fig:schematic}(a)].
Our results indicate that the significant contribution of $D_\mathrm{ion}$ to $D$ is explained by the excess osmotic pressure from the Donnan equilibrium \cite{donnan1911theory,adair1923donnan}.
These findings demonstrate the successful prediction of dynamic property from the static properties of charged gels, suggesting broader applicability of this relationship to charged gels in good solvents.

\textit{Theoretical derivation of collective diffusion coefficient in charged gels}.---We derive the contribution of $D_\mathrm{ion}$ to $D$, using a similar procedure as that for $D_\mathrm{mix}$ and $D_\mathrm{el}$ in neutral gels \cite{fujiyabu2018three,fujiyabu2019shear,kim2020mixing,fujiyabu2021temperature}. 
The osmotic bulk modulus $K \equiv c (\partial \Pi/\partial c)$ in Eq.~(\ref{eq:THB}) is quantified by total osmotic pressure $\Pi$, where $c$ is the polymer mass concentrations.
For charged gels, $\Pi$ is considered as the sum of polymer-solvent mixing $\Pi_\mathrm{mix}$, elastic $\Pi_\mathrm{el}$, and ionic $\Pi_\mathrm{ion}$ contributions \cite{flory1953principles,flory1943statistical1,flory1943statistical2,katchalsky1955polyelectrolyte,katchalsky1951equation,treloar1973elasticity,duvsek1993responsive} as
$\Pi = \Pi_\mathrm{mix} + \Pi_\mathrm{el}+ \Pi_\mathrm{ion}$,
where $\Pi_\mathrm{mix} \sim c^{3\nu/(3\nu-1)}$ \cite{yasuda2020universal,sakumichi2022semidilute} and $\Pi_\mathrm{el} = -G_0(c/c_0)^{1/3}$ \cite{james1949simple,horkay2000osmotic}.
Here, $\nu \approx 0.588$ is the excluded volume parameter \cite{flory1953principles}, and $G_0$ and $c_0$ is the shear modulus and polymer mass concentration, respectively, at the as-prepared state.
Further, the friction coefficient per unit volume $f$ in Eq.~(\ref{eq:THB}) follows $f \sim c^{3/2}$ \cite{fujiyabu2017permeation,fujiyabu2019shear,kim2020mixing}.
Notably, $\Pi_\mathrm{mix}$ and $f$ depend only on $c$ at a constant $T$ within the same polymer-solvent system, regardless of the polymer network structures \cite{yasuda2020universal,fujiyabu2017permeation,fujiyabu2019shear,kim2020mixing}.
Assuming same dependences of $\Pi_\mathrm{mix}$, $\Pi_\mathrm{el}$, and $f$ in charged gels, we can express each contribution to $D$ using Eq.~(\ref{eq:THB}) as
\begin{equation}
D_\mathrm{mix} = \frac{3\nu}{3\nu-1}
\frac{\Pi_\mathrm{mix}}{f},
\quad
D_\mathrm{el} = \frac{G}{f},
\label{eq:Dmix}
\end{equation}
and
\begin{equation}
D_\mathrm{ion} = \frac{c}{f} \frac{\partial \Pi_\mathrm{ion}}{\partial c}.
\label{eq:Dion}
\end{equation}
The $D_\mathrm{ion}$ originates from $\Pi_\mathrm{ion}$ produced by immobilized charge in polymer networks.
Notably, Eqs.~(\ref{eq:Dmix}) have been validated for neutral gels at as-prepared states \cite{fujiyabu2018three,fujiyabu2019shear,fujiyabu2021temperature} and swollen states \cite{kim2020mixing}.

We quantified $\Pi_\mathrm{ion}$ using the Donnan equilibrium \cite{donnan1911theory,ricka1984swelling,adair1923donnan}. 
For a polymer network with immobilized charge molar concentration $c_\mathrm{im}$ in a solution containing monovalent ions, the molar concentrations of mobile cations and anions inside the gel ($c^\mathrm{+}$ and $c^\mathrm{-}$) and in the outer salt solution ($c_\mathrm{out}^{+}$ and $c_\mathrm{out}^{-}$) satisfy the following electric neutrality conditions: 
(i) $c_\mathrm{im} + c^\mathrm{+} = c^\mathrm{-}$ for the inner gel, 
(ii) $c_\mathrm{out} \equiv c_\mathrm{out}^{+} = c_\mathrm{out}^{-}$ for the outer salt solution, and
(iii) $c^\mathrm{+} c^\mathrm{-}=(c_\mathrm{out})^2$ between inner gel and outer salt solution.
Combining the conditions (i) and (iii) yields
$c^\mathrm{\pm} = \frac{1}{2}c_\mathrm{im} \left(\sqrt{1 + 4\alpha^2} \mp 1 \right)$,
where $\alpha \equiv c_\mathrm{out}/c_\mathrm{im}$ ($\geq 0$) is the molar concentration ratio between the mobile ion in the outer salt solution and the immobilized charge in the polymer network.
Hence, the osmotic pressure due to the immobilized charge is
\begin{equation}
\Pi_\mathrm{ion} = RT(c^\mathrm{+} + c^\mathrm{-} - 2c_\mathrm{out}) = c_\mathrm{im} RT \left( \sqrt{1 + 4\alpha^2} - 2\alpha \right).
\label{eq:Pion}
\end{equation}
Substituting Eq.~(\ref{eq:Pion}) into Eq.~(\ref{eq:Dion}) gives
\begin{equation}
D_\mathrm{ion} = \frac{c_\mathrm{im} RT}{f\sqrt{1+4\alpha^2}}.
\label{eq:Dion2}
\end{equation}
According to Eq.~(\ref{eq:Dion2}), $\alpha$ regulates the efficiency of $D_\mathrm{ion}$, with $D_\mathrm{ion} =  c_\mathrm{im}RT/f$ in the charged gel limit ($\alpha \ll 1$) and $D_\mathrm{ion} = 0$ in the neutral gel limit ($\alpha \gg 1$).

\textit{Materials and methods}.---As a model system to investigate charged gels, we used a tetra-arm poly(ethylene glycol) (PEG) hydrogel, synthesized via the \textit{AB}-type cross-end coupling of two prepolymers (tetra-arm PEG) of equal size. 
Each end of the tetra-arm PEG was modified with a mutually reactive amine (tetra-PEG-NH$_{2}$) and succinimidyl ester (tetra-PEG-OSu) (NOF Co., Tokyo, Japan). 
All other reagents were purchased from WAKO Pure Chemicals (Osaka, Japan). 
All materials were used without further purification. 
We dissolved tetra-PEG-NH$_2$ and tetra-PEG-OSu (molar mass $11$ kg$/$mol) in phosphate buffer ($p\mathrm{H}= 7.0$, $60$ mM) to achieve a concentration of $c=60$ g$/$L. 

To fabricate charged gels, we mixed two solutions at stoichiometrically imbalanced ratios of $s = 0.325$, $0.350$, $0.375$, and $0.400$ (Table~\ref{tab:condition}), where $s$ is the molar fraction of minor tetra-PEG polymers ([tetra-PEG-OSu]) to total tetra-PEG polymers ([tetra-PEG-NH$_2$] $+$ [tetra-PEG-OSu]).
An excess tetra-PEG-NH$_{2}$ ($s > 0.5$) allowed some amine groups to remain partly unreacted \cite{yoshikawa2019connectivity}, serving as the immobilized cations in the polymer networks [see Fig.~\ref{fig:schematic}(b)].
We kept each sample in an enclosed space to maintain humid conditions at room temperature ($T \approx 298 K$) for reaction completion.

We immersed each charged gel into the outer salt (HCl $+$ NaCl) solutions of $p\mathrm{H}=4.0$ and $c_\mathrm{out} = 0.1$--$30$ mM (Table~\ref{tab:condition}) to completely protonate immobilized cations and to tune the parameter $\alpha \equiv c_\mathrm{out}/c_\mathrm{im}$ in Eq.~(\ref{eq:Dion2}).
We replaced the outer salt solution twice every $20$ hours to achieve equilibrium swollen state. 
The Henderson-Hasselbalch equation \cite{henderson1908concerning,hasselbalch1917calculation} determined the dissociation constant of immobilized residues as $K_{a} \approx 1.0 \times 10^{-10}$ M (see Supplemental Material, Sec.~S1 \cite{supplement}), confirming complete protonation of amines in all experimental conditions because $K_{a}$ was smaller than [H$^{+}$] $= 10^{-4}$ M.

We measured the diameter $d$ of each gel at equilibrium in the outer salt solutions, using an optical microscope (M165 C; Leica, Wetzlar, Germany) to calculate the volume swelling ratio $Q = (d/d_{0})^{3}$ (initial diameter $d_{0} = 1.04$ mm), determining $c=c_0/Q$ and $c_\mathrm{im}=c_\mathrm{im}^{0}(c/c_{0})$.
To ensure accurate $c_0$ and $c_\mathrm{im}^{0}$ values at each $s$, we assessed the sol fraction eluted out from the polymer network during the swelling process using the Bethe approximation \cite{supplement,macosko1976new,miller1976new} (Further details are described in Supplemental Material, Sec.~S2 \cite{supplement}).

\begin{table}[t!]
\caption{Synthetic conditions of charged gels ($s$, $c_0$, and $c_\mathrm{im}^0$) and experimental conditions for immersing charged gels in outer salt solutions ($c_\mathrm{out}$).
We evaluated $\alpha\equiv c_\mathrm{out}/c_\mathrm{im}$ and measured $G_{0}$ and $Q$ for each gel sample.
}
\label{tab:condition}
\begin{ruledtabular}
\begin{tabular}{ccccccc}
$s$ & $c_{0}$\,(g$/$L) & $c_\mathrm{im}^{0}$\,(mM) & $c_\mathrm{out}$\,(mM) & $c_\mathrm{out}/c_\mathrm{im}$ &  $G_{0}$\,(kPa) & $Q$ \\
\hline
\multirow{4}{*}{$0.400$} & \multirow{4}{*}{$57.2$} & \multirow{4}{*}{$3.4$} & $0.10$ & $0.079$ &  \multirow{4}{*}{$5.93$} & $2.7$ \\
& & & $3.0$ & $1.7$ & & $1.9$ \\
& & & $10.0$ & $5.3$ & & $1.8$ \\
& & & $30.0$ & $16.8$ & & $1.9$ \\
\hline
\multirow{4}{*}{$0.375$} & \multirow{4}{*}{$56.5$} & \multirow{4}{*}{$4.0$} & $0.10$ & $0.095$ & \multirow{4}{*}{$4.79$} & $3.8$ \\
& & & $1.1$ & $0.715$ & & $2.6$ \\
& & & $10.0$ & $5.0$ & & $2.0$ \\
& & & $30.0$ & $15.0$ & & $2.0$ \\
\hline
\multirow{4}{*}{$0.350$} & \multirow{4}{*}{$54.7$} & \multirow{4}{*}{$4.5$} & $0.10$ & $0.11$ & \multirow{4}{*}{$3.43$} & $4.8$ \\
& & & $1.1$ & $0.73$ & & $3.0$ \\
& & & $10.0$ & $4.9$ & & $2.2$ \\
& & & $30.0$ & $14.7$ & & $2.2$ \\
\hline
\multirow{4}{*}{$0.325$} & \multirow{4}{*}{$52.5$} & \multirow{4}{*}{$4.8$} & $0.10$ & $0.14$ & \multirow{4}{*}{$2.23$} & $6.7$ \\
& & & $0.30$ & $0.33$ & & $5.3$ \\
& & & $10.0$ & $6.5$ & & $3.1$ \\
& & & $30.0$ & $15.0$ & & $2.4$ \\
\end{tabular} 
\end{ruledtabular}
\end{table}

We measured $D$ of charged gels via DLS (ALV/CGS-3 compact goniometer, Langen, Germany) in the same way as Refs.~\cite{fujiyabu2018three,fujiyabu2019shear,kim2020mixing,fujiyabu2021temperature}.
Each gel sample was fabricated in a scattering glass tube (disposable culture tube 9830-1007 with an inner diameter of $8.4$ mm; IWAKI, Japan) and was swollen in the outer salt solution.
We carefully selected $d_{0}$ of each gel to prevent contact between a swollen gel and an inner wall of the glass tube.
Once equilibrium was reached, we measured the scattering light intensity $I(t)$ at time $t$ over $600$ s at a scattering wavelength $\lambda = 632.8$ nm and a scattering angle $\theta = \pi/2$, to evaluate the autocorrelation function \cite{johnson1994laser,berne2000dynamic} $g^{(2)}(\tau) \equiv \langle I(0)I(\tau)\rangle / \langle I(0) \rangle^{2}$ for delay time $\tau\approx0.01$--$0.1$ ms, corresponding to the thermal fluctuation of a polymer network \cite{li2017probe,fujiyabu2018three,fujiyabu2019shear,kim2020mixing,fujiyabu2021temperature,ohira2018dynamics} (results are provided in Supplemental Material, Sec.~S3 \cite{supplement}).
Here, $\langle \rangle$ denotes the time-average.
We fitted the autocorrelation functions using a stretched exponential function $g^{2}(\tau) - 1 = A \exp [-2(\tau/\tau_{g})^{B}] + \epsilon$, where $A$ is the initial amplitude, $B$ is the stretched exponent, and $\epsilon$ is the time-independent background.
Using the partial heterodyne model \cite{joosten1991dynamic,shibayama2002gel}, we evaluate $D = (1+\sqrt{1+A})/(q^{2}\tau_{g})$, where $q = (4\pi n/\lambda)\sin(\theta/2) \approx 0.0187$ nm$^{-1}$ is the scattering vector with the refractive index $n \approx 1.333$  of the aqueous solvents.

We measured the (equilibrium) shear modulus of gels at the equilibrium swollen state $G$ for each $c_\mathrm{out}$ using a rheometer (MCR302; Anton Paar, Graz, Austria) with a $25$-mm parallel plate geometry.
The gels, shaped into discs with a height $1$ mm and a diameter $35$ mm, were equilibrated in outer salt solutions with varying $c_\mathrm{out}$ (Table~\ref{tab:condition}). 
We measured the storage modulus $G'$ and loss modulus $G''$ at $T=298$ K with angular frequency $\omega = 0.63$--$63$ rad$/$s and shear strain $\gamma = 1.0$\%, as well as $\omega = 6.3$ rad$/$s and $\gamma = 0.1$\% --$10$\%. 
For all samples, $G'$ was independent of $\gamma$ and $\omega$, and was much larger than $G''$ (Supplemental Material, Sec.~S4 \cite{supplement}).
Hence, we selected $G'$ measured at $\omega = 6.3$ rad$/$s and $\gamma = 1.0$\% as the (equilibrium) shear modulus $G$.
Using $G = G_0(c/c_0)^{1/3}$ for swollen gels \cite{james1949simple,horkay2000osmotic} (see Supplemental Material, Sec.~S5 \cite{supplement}), we evaluated $G_0$ at the as-prepared state as $G_0 = G(c_0/c)^{1/3}$ for each $s$ (Table~\ref{tab:condition}).

\begin{figure}[t!]
\centering
\includegraphics[width=\linewidth]{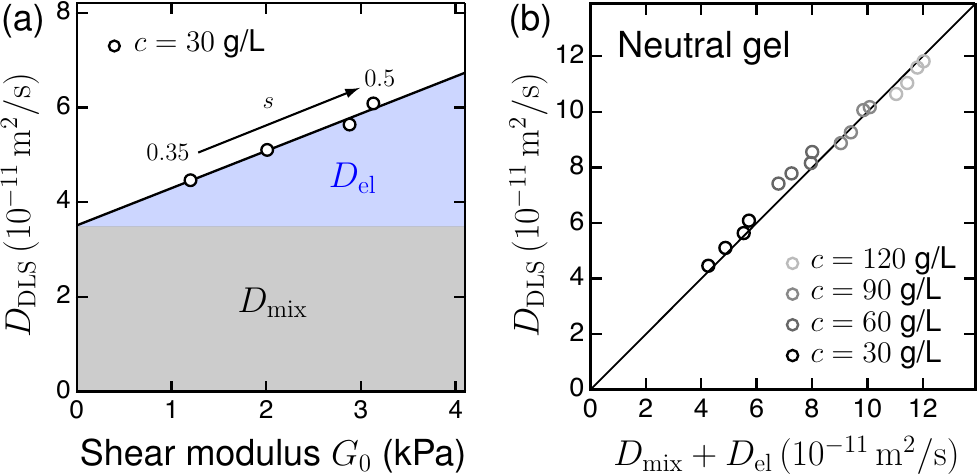}
\caption{
Collective diffusion coefficient for neutral gels.
(a) Typical result of shear modulus ($G_0$) dependence of the collective diffusion coefficient and its components ($D_\mathrm{mix}$ and $D_\mathrm{el}$) for neutral gels at the as-prepared state.
We tune $s$ at the constant $c$ so that $D_\mathrm{el}$ can be controlled independently. 
(b) Comparison between the collective diffusion coefficients measured via DLS ($D_\mathrm{DLS}$) and calculated by Eqs.~(\ref{eq:Dmix}) ($D_\mathrm{mix} + D_\mathrm{el}$) for neutral gels, using $D_\mathrm{mix} = 2.2 \times 10^{-12}c^{0.8}$ and $D_\mathrm{el}=G_{0}/(8.0\times10^{11}c^{1.5})$.
The solid line represents $D_\mathrm{DLS} = D_\mathrm{mix} + D_\mathrm{el}$.
In (a) and (b), the data of neutral gels for $c=30, 60, 90$, and $120$~g$/$L with the molar mass $20$~kg$/$mol at $s=0.35, 0.4, 0.45$, and $0.5$ are taken from Ref.~\cite{fujiyabu2021temperature}.
}
\label{fig:neutral}
\end{figure}

\textit{Estimating $D_\mathrm{mix}$ and $D_\mathrm{el}$ in neutral gels}.---Before considering charged gels, we briefly revisit the decomposition of mixing ($D_\mathrm{mix}$) and elastic ($D_\mathrm{el}$) contributions to $D$ in a model neutral gels, based on our previous studies \cite{fujiyabu2018three,fujiyabu2019shear,kim2020mixing,fujiyabu2021temperature}.
Figure~\ref{fig:neutral}(a) shows the collective diffusion coefficient measured via DLS $D_\mathrm{DLS}$ in neutral gels at the as-prepared state, plotted against $G_0$ for various $s$.
The observed linear relationship between $D_\mathrm{DLS}$ and $G_0$ is consistent with Eqs.~(\ref{eq:Dmix}). 
Notably, $D_\mathrm{mix}$ is independent of $s$, because $\Pi_\mathrm{mix}$ and $f$ is independent of $s$ \cite{yasuda2020universal,fujiyabu2017permeation,fujiyabu2019shear,kim2020mixing}.
Thus, we can decompose $D_\mathrm{DLS}$ into the mixing contribution $D_\mathrm{mix}=\lim_{G\to0} D_\mathrm{DLS}$ and the elastic contribution $D_\mathrm{el}=D_\mathrm{DLS}-D_\mathrm{mix}$ [see Fig.~\ref{fig:neutral}(a)]. 

\begin{figure}[t!]
\centering
\includegraphics[width=\linewidth]{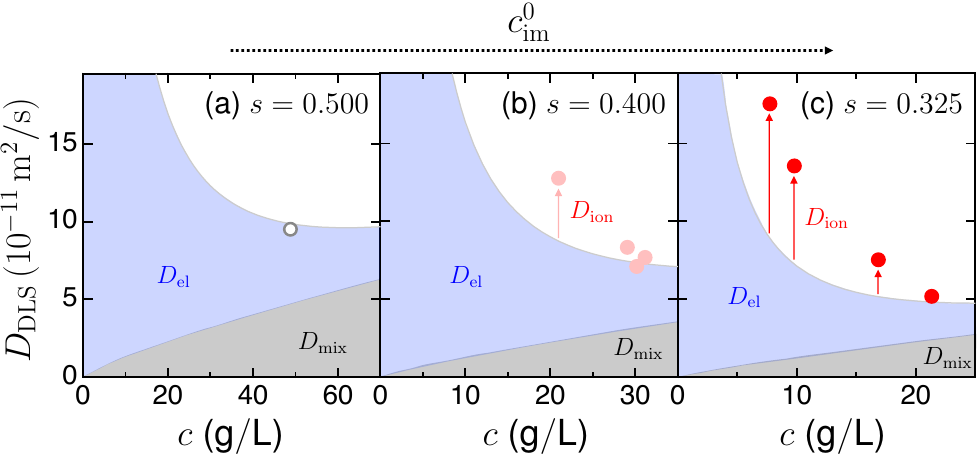}
\caption{
Polymer mass concentration ($c$) dependence of the collective diffusion coefficient and its components ($D_\mathrm{mix}$ and $D_\mathrm{el}$) for neutral gels at (a) $s = 0.500$ and charged gels for varying $c_\mathrm{out}$ at (b) $s = 0.400$ and (c) $s = 0.325$.
The $c_\mathrm{im}^0$ increases as $s$ decreases (black dotted arrow).
For charged gels, $D_\mathrm{mix} = 2.2 \times 10^{-12}c^{0.8}$ and $D_\mathrm{el}=G_{0}(c/c_{0})^{1/3}/(8.0\times10^{11}c^{1.5})$ were assumed to be equal to neutral gels.
The results at $s=0.375$ and $s=0.350$ are shown in Supplemental Material, Sec.~S6 \cite{supplement}.
}
\label{fig:D-c}
\end{figure}

Through the decomposition of $D_\mathrm{mix}$ and $D_\mathrm{el}$, we obtain the empirical formulas $D_\mathrm{mix} = 2.2 \times 10^{-12}c^{0.8}$ and $f = 8.0\times10^{11}c^{1.5}$ for neutral gels (the detailed results are provided in Supplemental Material, Sec.~S7), corroborating the scaling predictions $f \sim c^{1.5}$ \cite{tokita1991friction} and $\Pi_\mathrm{mix} \sim c^{2.3}$ \cite{des1975lagrangian,de1979scaling} (for gels in a good solvent).
These formulas are applicable to neutral gels at the as-prepared and swollen states, as confirmed in our previous study \cite{kim2020mixing}.
Using values of $c$ and $G$, we can accurately calculate $D$ for a neutral gel.
Figure~\ref{fig:neutral}(b) demonstrates that the collective diffusion coefficient of neutral gels measured via DLS ($D_\mathrm{DLS}$) shows excellent agreement with Eqs.~(\ref{eq:Dmix}), using empirical formulas of $D_\mathrm{mix}$ and $f$.

\textit{Estimating $D_\mathrm{ion}$ in charged gels}.---
We expand our analysis to electrically charged gels and examine Eq.~(\ref{eq:D-ion-gel}), by hypothesizing that the formulas of $D_\mathrm{mix}$ and $D_\mathrm{el}$, as established for neutral gels, remain applicable.
Figure~\ref{fig:D-c} shows $D_\mathrm{DLS}$ for swollen neutral and charged gels (equilibrated in outer salt solutions with varying $c_\mathrm{out}$), as measured via DLS.
For varying $s$ samples, a decrease in $s$ corresponded with an increase in $c_\mathrm{im}^0$ and a decrease in $G_0$.
This resulted in an increase in $Q$ to be a decrease in $c$ ($= c_{0}/Q$) in outer salt solutions (Table~\ref{tab:condition}).
The decrease in $G_0$ is attributed to a decrease in elastically effective subchains in polymer networks \cite{sakai2008design}.
Moreover, for a fixed $s$, a decrease in $c_\mathrm{out}$ in outer salt solutions led to a lower $\alpha$ ($\equiv c_\mathrm{out}/c_\mathrm{im}$), resulting in increased $Q$ to be decreased $c$ ($= c_{0}/Q$) (Table~\ref{tab:condition}). 

For neutral gels [Fig.~\ref{fig:D-c}(a)], $D_\mathrm{DLS}$ agrees with $D_\mathrm{mix} + D_\mathrm{el}$, which is independent of $c_\mathrm{out}$ (see Supplemental Material, Sec.~S8 \cite{supplement} for further details).
For charged gels with each $s$ [Fig.~\ref{fig:D-c}(b) and (c)], we calculate $D_\mathrm{mix}$ (gray region) and $D_\mathrm{el}$ (blue region), using the same empirical formulas for neutral gels with all measurable $G_0$ and $Q$.
Here, we experimentally confirmed that $G=G_0(c/c_0)^{1/3}$ holds for charged gels (see Supplemental Material, Sec.~S5 \cite{supplement}), indicating that the immobilized charge does not significantly affect the elastic modulus.
At high $c_\mathrm{out}$ ($\alpha \gg 1$), $D_\mathrm{DLS}$ approaches the neutral gel limit ($D_\mathrm{mix} + D_\mathrm{el}$), because $D_\mathrm{ion} \to 0$ in Eq.~(\ref{eq:Dion2}).
In contrast, at low $c_\mathrm{out}$ ($\alpha \ll 1$), $D_\mathrm{DLS}$ deviates from the neutral gel limit ($D = D_\mathrm{mix} + D_\mathrm{el}$), indicating a significant $D_\mathrm{ion}$ to $D$ (indicated by red arrows), which contributed up to approximately half of $D$ at its maximum.

\begin{figure}[t!]
\centering
\includegraphics[width=\linewidth]{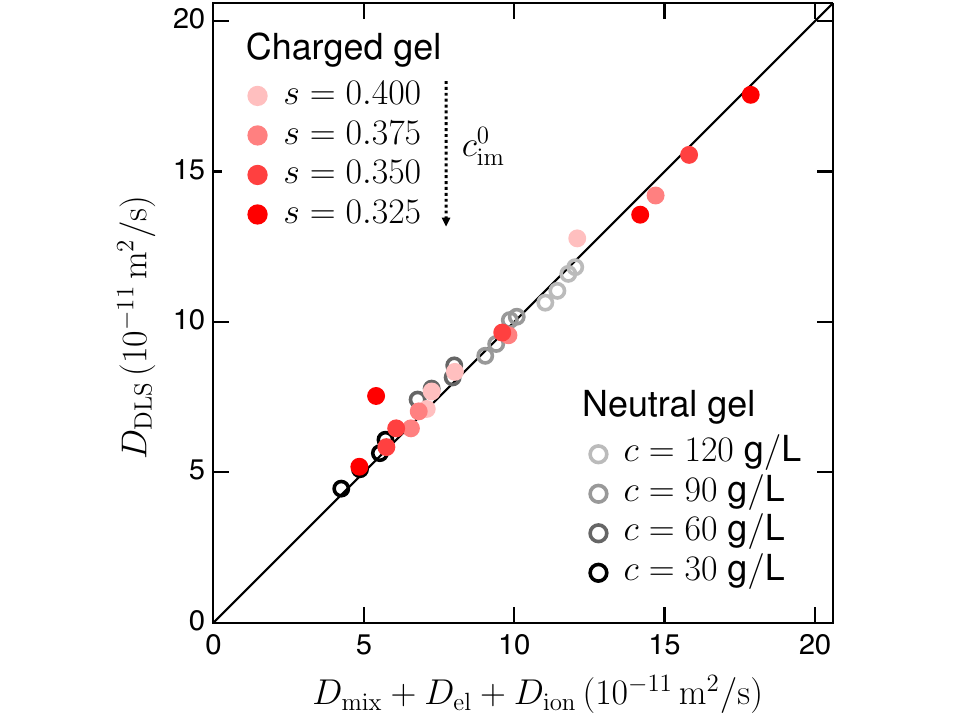}
\caption{
Comparison between the collective diffusion coefficients measured via DLS ($D_\mathrm{DLS}$) and calculated by Eqs.~(\ref{eq:Dmix}) and (\ref{eq:Dion}) ($D_\mathrm{mix} + D_\mathrm{el} + D_\mathrm{ion}$) for charged gels.
Here, $D_\mathrm{mix}$ and $D_\mathrm{el}$ were assumed to be equal in neutral gels: $D_\mathrm{mix} = 2.2 \times 10^{-12}c^{0.8}$ and $D_\mathrm{el}=G_{0}(c/c_{0})^{1/3}/(8.0\times10^{11}c^{1.5})$, and $D_\mathrm{ion}$ was calculated by Eq.~(\ref{eq:Dion2}) with each $c_\mathrm{im}$ and $\alpha$.
The solid line represents $D_\mathrm{DLS} = D_\mathrm{mix} + D_\mathrm{el} + D_\mathrm{ion}$.
}
\label{fig:D-comparison}
\end{figure}

We show that $D$ of charged gels can be accurately described by simply adding $D_\mathrm{ion}$ in Eq.~(\ref{eq:Dion2}).
Figure~\ref{fig:D-comparison} demonstrates that the collective diffusion coefficient of charged gels measured via DLS ($D_\mathrm{DLS}$) shows excellent agreement with the predictions of Eqs.~(\ref{eq:Dmix}) and (\ref{eq:Dion}), using the same empirical formulas of $D_\mathrm{mix}$ and $f$ as in neutral, and applying Eq.~(\ref{eq:Dion2}) with each $c_\mathrm{im}$ and $\alpha$.
These findings suggest that Eq.~(\ref{eq:THB}) can be extended to include charged gels by simply adding $D_\mathrm{ion}$ without evident cross-correlations.
Notably, Eq.~(\ref{eq:Dion2}) is applicable in weakly charged gels and is less effective in polyelectrolyte gels with high $c_\mathrm{im}$, due to counterion condensation \cite{tang2020swelling,manning1969limiting}.

\textit{Concluding remarks}.---
We developed model charged gels [Fig.~\ref{fig:schematic} and Table~\ref{tab:condition}] with comparable $D_\mathrm{mix}, D_\mathrm{el},$ and $D_\mathrm{ion}$ and measured their collective diffusion coefficient $D$ via DLS.
Based on the same decomposition of $D_\mathrm{mix}$ and $D_\mathrm{el}$ to $D$ in neutral gels [Figs.~\ref{fig:neutral} and \ref{fig:D-c}], our results indicate that $D$ in the charged gel is governed by Eq.~(\ref{eq:THB}), where $D_\mathrm{ion}$ emerges as an additive component [Fig.~\ref{fig:D-comparison}].
The significant ionic contribution $D_\mathrm{ion}$ to $D$ is quantitatively described through the ionic osmotic pressure $\Pi_\mathrm{ion}$ [Eqs.~(\ref{eq:Pion}) and (\ref{eq:Dion2})], which originates from the Donnan equilibrium.

Our findings demonstrate the successful prediction of the dynamic behaviors from the static properties of charged gels equilibrated in a good solvent.
This aspect is particularly relevant considering the dynamics in gel-like biological systems such as extracellular matrices, which are predominantly negatively charged \cite{theocharis2016extracellular} and influenced by the Donnan effect.
Therefore, an in-depth understanding of the dynamics in charged gels will enhance our comprehension of the behaviors of biological systems.

\begin{acknowledgments}
This work was supported by the Japan Society for the Promotion of Science (JSPS) through the Grants-in-Aid for JSPS 
Research Fellows Grant No.~19J22561 to T.F. and No.~202214177 to T.Y.,
Scientific Research (B) Grant No.~22H01187 to N.S.,
Scientific Research (A) Grant No.~21H04688 to T.S.,
Transformative Research Area Grant No.~20H05733 to T.S.,
and MEXT Program Grant No.~JPMXP1122714694 to T.S.
This work was also supported by JST through CREST Grant No.~JPMJCR1992 and Moon-shot R\&D Grant No.~1125941 to T.S.
\end{acknowledgments}

\bibliographystyle{apsrev}

\end{document}

\title{Supplemental Material for:
``Polymer Network Diffusion in Charged Gels''}

\author{Shoei~Sano}
\affiliation{Department of Bioengineering, The University of Tokyo, Hongo, Bunkyo-ku, Tokyo 113-8656, Japan.}
\author{Takashi~Yasuda}
\affiliation{Department of Bioengineering, The University of Tokyo, Hongo, Bunkyo-ku, Tokyo 113-8656, Japan.}
\author{Takeshi~Fujiyabu}
\affiliation{Department of Bioengineering, The University of Tokyo, Hongo, Bunkyo-ku, Tokyo 113-8656, Japan.}
\author{Naoyuki~Sakumichi}
\email[Correspondence should be addressed to N.~Sakumichi or T.~Sakai: ]
{sakumichi@gel.t.u-tokyo.ac.jp}
\affiliation{Department of Chemistry and Biotechnology, The University of Tokyo, Hongo, Bunkyo-ku, Tokyo 113-8656, Japan.}
\author{Takamasa~Sakai}
\email{sakai@gel.t.u-tokyo.ac.jp}
\affiliation{Department of Chemistry and Biotechnology, The University of Tokyo, Hongo, Bunkyo-ku, Tokyo 113-8656, Japan.}
\date{\today}

\maketitle
\noindent

%%%%%%%%%% Prefix a "S" to all equations, figures, tables and reset the counter %%%%%%%%%%
\setcounter{equation}{0}
\setcounter{figure}{0}
\setcounter{table}{0}
\setcounter{page}{1}
\makeatletter
\renewcommand{\theequation}{\thesection.\arabic{equation}}
\renewcommand{\theequation}{S\arabic{equation}}
\renewcommand{\thefigure}{S\arabic{figure}}
\renewcommand{\bibnumfmt}[1]{[S#1]}
\renewcommand{\citenumfont}[1]{S#1}
 \@addtoreset{equation}{section}
%%%%%%%%%% Prefix a "S" to all equations, figures, tables and reset the counter %%%%%%%%%%

\section{CALCULATION OF DISSOCIATION CONSTANT FOR THE IMMOBILIZED BASE}

\begin{figure}[h!]
\centering
\includegraphics[width=\linewidth]{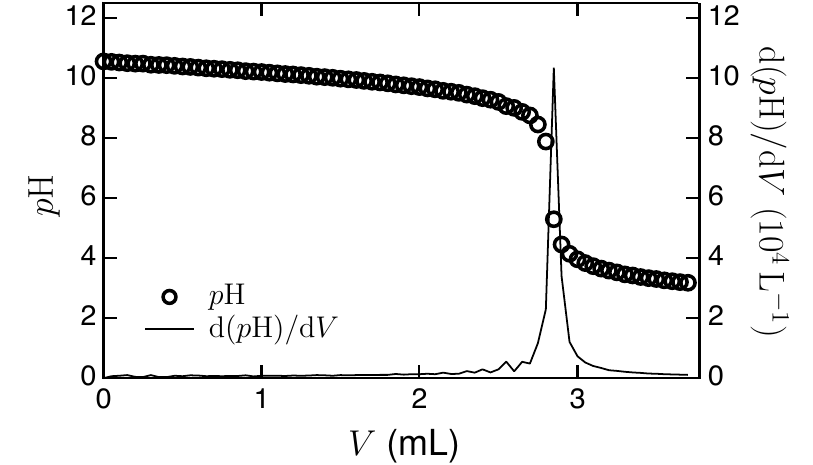}
\caption{Titration curve of tetra-PEG-NH$_{2}$ solutions with HCl solution of $10$ mM.
}
\label{fig:titration}
\end{figure}

We determine the dissociation constant ($\mathrm{K}_{a}$) of the amine group on the ends of the prepolymer (tetra-PEG-NH$_{2}$) using an automatic titration apparatus (COM-1700A; HIRANUMA, Ibaraki, Japan).
We added HCl solutions of $10$~mM to tetra-PEG-NH$_{2}$ solution for $c=10$~g$/$L at $T=298$ K to neutralize the amine groups (NH$_{3}^{+}$ to NH$_{2}$).
Figure~\ref{fig:titration} shows the decrease in \textit{p}H with increasing titer ($V$), indicating $\mathrm{d}(p\mathrm{H})/\mathrm{d}V$ shows a sharp peak at $V \approx 2.85$~mL.
Using the Henderson-Hasselbalch equation \cite{henderson1908concerning,hasselbalch1917calculation} for the point where [NH$_{3}^{+}$] $=$ [NH$_{2}$], we calculated $\mathrm{K}_{a}$ as
\begin{equation}
p\mathrm{K}_{a} = p\mathrm{H} + \log{\frac{[\mathrm{NH}_{3}^{+}]}{[\mathrm{NH}_2]}}.
\label{eq:Ka}
\end{equation}
Accordingly, we estimated $p\mathrm{K}_{a} \approx 10.0$, yielding $\mathrm{K}_{a} \approx 1.0 \times 10^{-10}$ M.

\section{EVALUATION OF SOL FRACTION THROUGH BETHE APPROXIMATION}

To accurately determine $c_0$ and $c_\mathrm{im}^{0}$ of polymer gels immersed in outer solutions, we evaluate the sol fraction eluting from the polymer network during the swelling process using the Bethe approximation \cite{macosko1976new,miller1976new}.
Considering the \textit{AB}-type end coupling between tetra-functional polymers (A4 and B4), the Bethe approximation gives the probability that one arm of each polymer does not connect to the infinite network [$P(F_\mathrm{A}^\mathrm{out})$ and $P(F_\mathrm{B}^\mathrm{out})$] as
\begin{equation}
P(F_\mathrm{A}^\mathrm{out})=p_\mathrm{A} P(F_\mathrm{B}^\mathrm{out})^3+1-p_\mathrm{A},
\label{eq:PFA}
\end{equation}
\begin{equation}
P(F_\mathrm{B}^\mathrm{out})=p_\mathrm{B} P(F_\mathrm{A}^\mathrm{out})^3+1-p_\mathrm{B},
\label{eq:PFB}
\end{equation}
where $p_\mathrm{A}$ and $p_\mathrm{B}$ are the probability of reacted A and B end groups, respectively.
Considering the stoichiometrically imbalanced mixing with a ratio [A4] : [B4] $= s$ : ($1-s$) for $s<0.5$, we have $p_\mathrm{A} = p_\mathrm{B}[(1-s)/s]$.
Assuming $p_\mathrm{A} \approx 0.92$ for incomplete reaction conversion, Eqs.~(\ref{eq:PFA}) and (\ref{eq:PFB}) give $P(F_A^\mathrm{out})$ and $P(F_B^\mathrm{out})$ at each $s$, leading to
\begin{equation}
c_{0}=c_0^\mathrm{(pre)}
\{ 
s[1-P(F_\mathrm{A}^\mathrm{out})^4]
+(1-s)[1-P(F_\mathrm{B}^\mathrm{out})^4]
\},
\label{eq:bethe}
\end{equation}
where $c_0^\mathrm{(pre)}$ is the polymer mass concentration at pre-immersion, and
\begin{equation}
c_\mathrm{im}^0=4c_0/M
\{
(1-s)[1-P(F_\mathrm{B}^\mathrm{out})^4]
-s[1-P(F_\mathrm{A}^\mathrm{out})^4]
\},
\label{eq:cim0}
\end{equation}
where $M$ is the molar mass of prepolymers. 

\begin{table}[b!]
\caption{Comparison of polymer mass concentration of gels immersed in outer solutions measured by weight change ($c_\mathrm{0,exp}$) and evaluated through the Bethe approximation ($c_0$) at each $s$.
}
\label{tab:bethe}
\begin{ruledtabular}
\begin{tabular}{ccccc}
s & $W_\mathrm{polymer}^\mathrm{(pre)}$ (mg) & $W_\mathrm{polymer}$ (mg) & $c_\mathrm{0,exp}$ (g$/$L) & $c_\mathrm{0}$ (g$/$L) \\
\hline
$0.400$ & $349$ & $328$ & $56.4$ & $57.2$ \\
$0.375$ & $356$ & $326$ & $55.0$ & $56.5$ \\
$0.350$ & $175$ & $157$ & $53.9$ & $54.7$ \\
$0.325$ & $172$ & $149$ & $52.1$ & $52.5$ \\
\end{tabular}
\end{ruledtabular}
\end{table}

To validate Eqs.~(\ref{eq:bethe}) and (\ref{eq:cim0}), we experimentally measured the polymer mass concentration immersed in outer solutions $c_\mathrm{0,exp}$ at each $s$ and compared with $c_0$ evaluated through the Bethe approximation [Eq.~(\ref{eq:bethe})].
We measured the weight of a gel at pre-immersion $W_{0}^\mathrm{(pre)}$ and immersed a gel in pure water for $48$ hours. 
This process was repeated twice to completely remove the sol fraction.
Subsequently, we dried the gel at $T = 298$ K for $48$ hours and measured the weight $W_\mathrm{polymer}$.
Here, the weight of polymers in the gel at pre-immersion is
$W_\mathrm{polymer}^\mathrm{(pre)} = W_{0}^\mathrm{(pre)} [c_{0}^\mathrm{(pre)}/(\rho_\mathrm{water} + c_{0}^\mathrm{(pre)}$)], 
where $\rho_\mathrm{water} \approx 997$ kg$/$m$^{3}$ is the density of water.
Using $W_\mathrm{polymer}$ and $W_\mathrm{polymer}^\mathrm{(pre)}$, we determine
$c_\mathrm{0,exp} = c_{0}^\mathrm{(pre)} (W_\mathrm{polymer}/W_\mathrm{polymer}^\mathrm{(pre)})$,
as listed in Table~\ref{tab:bethe}.
The measured $c_\mathrm{0,exp}$ is in good agreement with $c_{0}$, supporting the validity of Bethe approximation.

\section{AUTOCORRELATION FUNCTION MEASURED VIA DYNAMIC LIGHT SCATTERING}

\begin{figure}[t!]
\centering
\includegraphics[width=\linewidth]{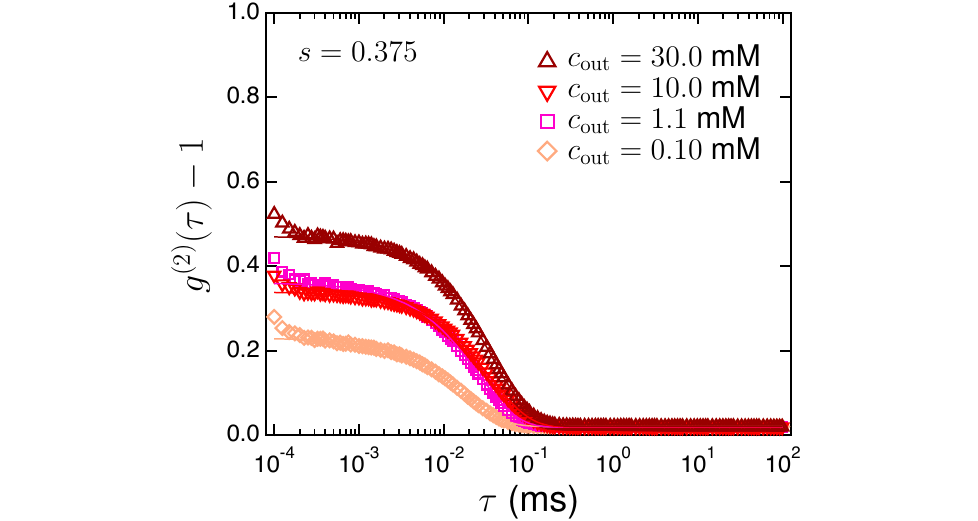}
\caption{
Autocorrelation functions [$g^{(2)}(\tau)-1$] of charged gel at $s = 0.375$ in outer salt solutions with varying $c_\mathrm{out} = 0.10$--$30$ mM measured via dynamic light scattering at $T = 298$ K.
Each solid curve is a least-squares fit of the stretched exponential function.
}
\label{fig:autocorrelation}
\end{figure}

\begin{table}[t!]
\caption{Least-square fitting parameters of stretched exponential function applied to autocorrelation functions in Fig.~\ref{fig:autocorrelation}.
}
\label{tab:fit}
\begin{ruledtabular}
\begin{tabular}{ccccc}
Outer salt solutions & A & B & $\tau_{g}$ (ms) & $\epsilon$ \\
\hline
$c_\mathrm{out}=30.0$ mM & $0.45$ & $0.94$ & $0.077$ & $0.021$ \\
$c_\mathrm{out}=10.0$ mM & $0.32$ & $0.94$ & $0.074$ & $0.020$ \\
$c_\mathrm{out}=1.1$ mM & $0.34$ & $0.93$ & $0.054$ & $0.020$ \\
$c_\mathrm{out}=0.10$ mM & $0.21$ & $0.90$ & $0.038$ & $0.021$ \\
\end{tabular} 
\end{ruledtabular}
\end{table}

Figure~\ref{fig:autocorrelation} shows typical results of the autocorrelation functions for charged gels in outer salt solutions measured via dynamic light scattering.
We fit each autocorrelation function using the stretched exponential function $g^{2}(\tau) - 1 = A \exp [-2(\tau/\tau_{g})^{B}] + \epsilon$, with the fitting parameters listed in Table~\ref{tab:fit}.
We evaluate $D = (1+\sqrt{1+A})/(q^{2}\tau_{g})$, using partial heterodyne model \cite{joosten1991dynamic,shibayama2002gel}.

\begin{figure}[t!]
\centering
\includegraphics[width=\linewidth]{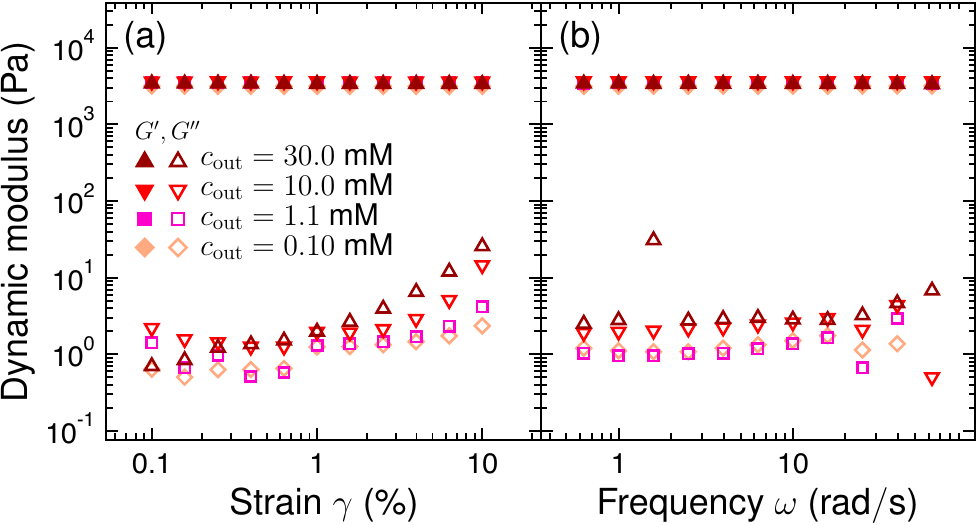}
\caption{Storage modulus $G'$ and loss modulus $G''$ of charged gels at $s = 0.375$ in outer salt solutions with varying $c_\mathrm{out}=0.10$--$30$ mM measured via rheometry at $298$ K.
(a) $\gamma$-dependence at $\omega = 6.3$ rad$/$s and (b) $\omega$-dependence at $\gamma = 1.0 \%$.
}
\label{fig:G1}
\end{figure}

\section{DYNACMIC VISCOELATICITY MEASUREMENTS OF CHARGED GEL IMMERSED IN OUTER SALT SOLUTIONS}

Figure~\ref{fig:G1} shows typical results of strain ($\gamma$) and frequency ($\omega$) dependencies of storage ($G'$) and loss ($G''$) moduli in charged gels in outer salt solutions measured via rheometry.
The $G'(\omega)$ is within the linear response region and is independent of the frequency ($\omega$) below $100$ rad$/$s. 
We regard $G'(\omega)$ at $1$ rad$/$s as the shear modulus $G$, because the equilibrium shear modulus is given by $G=\lim_{\omega \to 0} G'(\omega)$.

\begin{figure}[t!]
\centering
\includegraphics[width=1\linewidth]{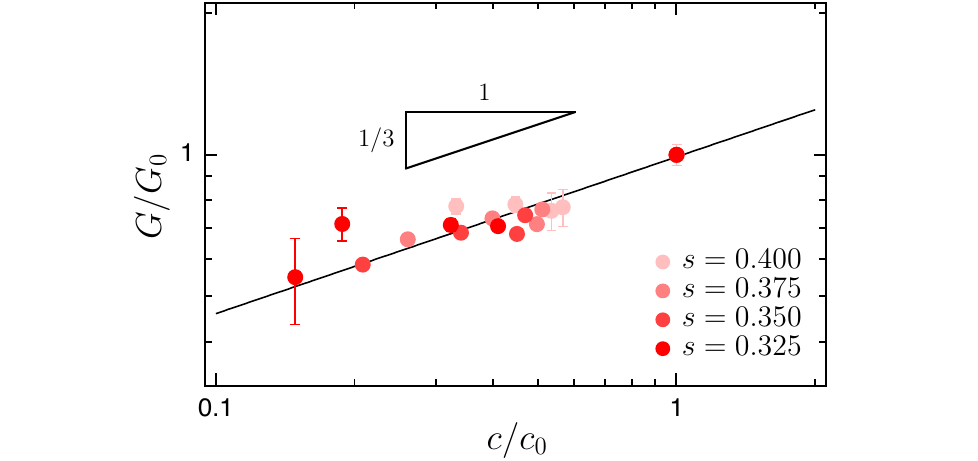}
\caption{Dependence of $G/G_{0}$ on $c/c_{0}$ in charged gels with varying $s$.
The error bars represent the experimental errors.
The solid line represents $G/G_{0}=(c/c_{0})^{1/3}$.
}
\label{fig:G2}
\end{figure}

\section{DECREASE OF SHEAR MODULUS $G$ DURING SWELLING IN CHARGED GELS}

We demonstrate the decrease of shear modulus $G=G_0(c/c_0)^{1/3}$ during the swelling process for charged gels.
This nonlinear relationship is consistent with models proposed by Flory \cite{flory1953principles} and Panyukov \cite{panyukov1990scaling}, which consider the competition of dilution and pre-extension of the polymer network due to swelling.
Figure~\ref{fig:G2} shows the normalized elastic modulus $G/G_{0}$ as a function of $c_{0}/c$.
All results are consistent with $G/G_{0}=(c/c_{0})^{1/3}$, indicating that the immobilized charge does not significantly influence the elastic modulus in charged gels.

\section{EMPIRICAL FORMULAS OF $D_\mathrm{mix}$ AND $f$ DEMONSTRATED IN NEUTRAL GELS}

\begin{figure}[t!]
\centering
\includegraphics[width=1\linewidth]{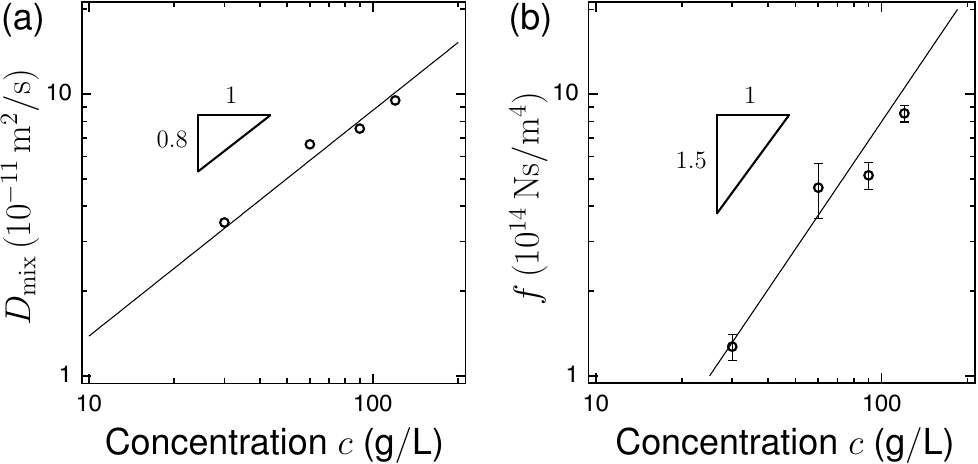}
\caption{
Polymer mass concentration ($c$) dependences of $D_\mathrm{mix}$ and $f$ evaluated through the decomposition of $D$ in neutral gels [Fig.~2(a) in the main text].
The error bars represent the standard error.
Each solid line represents the empirical formulas $D_\mathrm{mix} = 2.2 \times 10^{-12}c^{0.8}$ and $f=8.0\times10^{11}c^{1.5}$.
}
\label{fig:Dmix}
\end{figure}

We confirm that the empirical formulas of $D_\mathrm{mix} = 2.2 \times 10^{-12}c^{0.8}$ and $f=8.0\times10^{11}c^{1.5}$ used in the main text are consistent with the values of $D_\mathrm{mix}$ and $f(=G/D_\mathrm{el})$ evaluated through the decomposition of $D$ in neutral gels measured via dynamic light scattering.
Figures~\ref{fig:Dmix}(a) and (b) show $c$-dependence of $D_\mathrm{mix}$ and $f$, respectively, showing excellent agreements with the empirical formulas.\\

\begin{figure}[t!]
\centering
\includegraphics[width=1\linewidth]{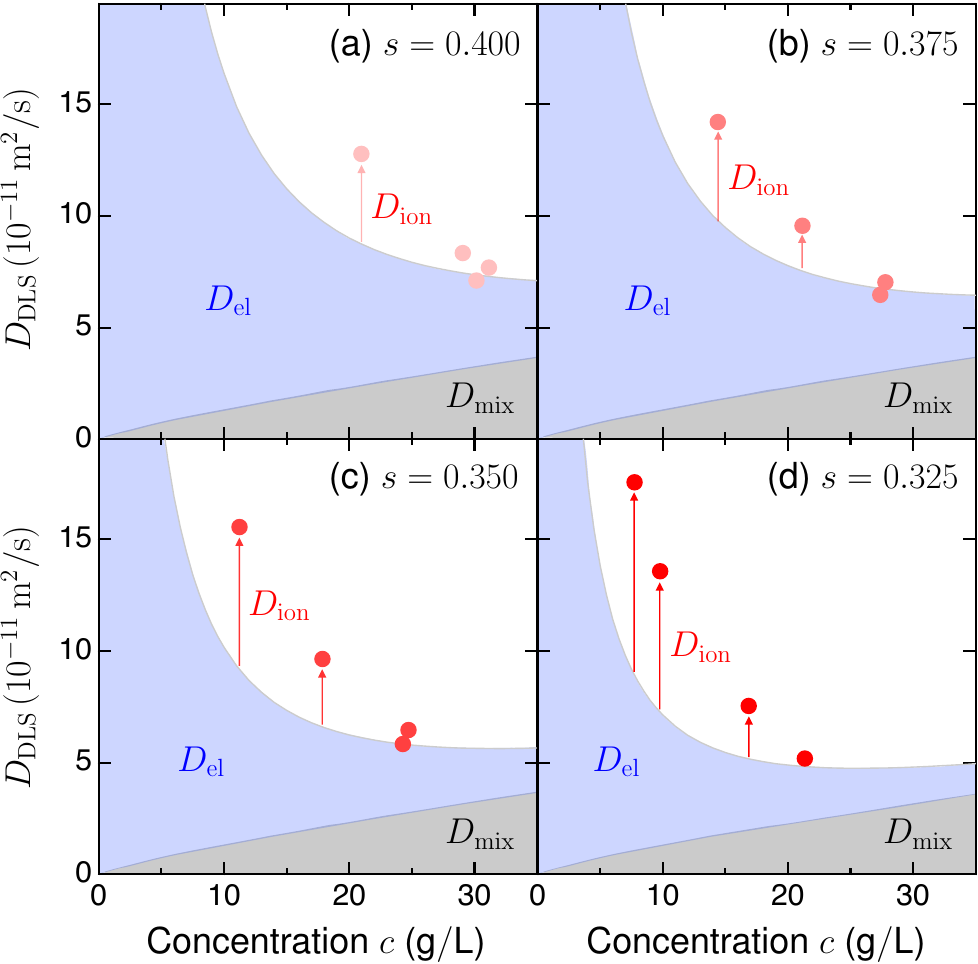}
\caption{
Polymer mass concentration dependences of collective diffusion coefficient and its components ($D_\mathrm{mix}$ and $D_\mathrm{el}$) in charged gels at (a) $s = 0.400$, (b) $s = 0.375$, (c) $s=0.350$, and (d) $s=0.325$ for varying $c_\mathrm{out}$. 
}
\label{fig:s}
\end{figure}

\begin{figure}[b!]
\centering
\includegraphics[width=1\linewidth]{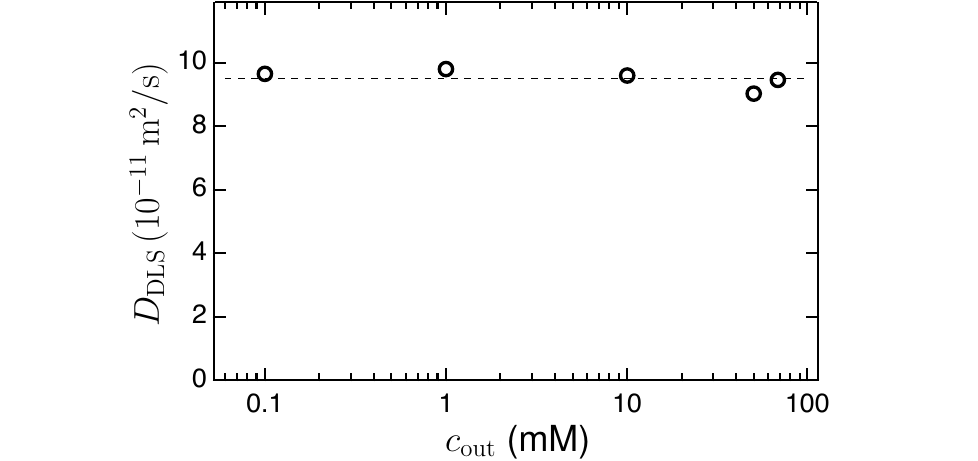}
\caption{
Dependence of $D_\mathrm{DLS}$ measured via dynamic light scattering on $c_\mathrm{out}$ in a neutral gel at $s=0.500$. 
The dotted line represents the average of $D_\mathrm{DLS}$.
}
\label{fig:cout}
\end{figure}

\section{POLYMER MASS CONCENTRATION DEPENDENCE OF $D$ IN CHARGED GELS AT EACH $s$}

Figure~\ref{fig:s} shows $c$-dependence of $D_\mathrm{DLS}$ measured via dynamic light scattering at each $s$, where $s=0.375$ and $s=0.350$ are not shown in the main text.
When $c_\mathrm{out}$ is sufficiently low to be $\alpha \gg 1$, $D_\mathrm{DLS}$ deviates from the neutral gel limit ($D = D_\mathrm{mix} + D_\mathrm{el}$) owing to the significant ionic contribution $D_\mathrm{ion}$ to $D$ (red arrows) at each $s$.

\section{EFFECT OF SALT CONCENTRATION ON COLLECTIVE DIFFUSION COEFFICIENT IN NEUTRAL GELS}

We investigate the effect of $c_\mathrm{out}$ on $D$ in a neutral gel.
Figure~\ref{fig:cout} demonstrates that $D_\mathrm{DLS}$ ($= D_\mathrm{mix} + D_\mathrm{el}$), as measured via dynamic light scattering, remains unaffected by $c_\mathrm{out}$ within the experimental range of this study ($c_\mathrm{out} = 0.1$--$30$ mM).
The data for $D_\mathrm{DLS}$ at $s=0.500$ presented in Fig.~3(a) in the main text is derived from the average of $D_\mathrm{DLS}$ shown in Fig.~\ref{fig:cout}.